\documentclass[aps,prl,reprint,showpacs]{revtex4-1}

\usepackage{graphicx}
\usepackage{amsfonts}
\usepackage{amsmath}
\usepackage[utf8]{inputenc}
\usepackage{hyperref}
\usepackage{varioref}
\usepackage{ulem}

\labelformat{equation}{\textup{(#1)}}

\newcommand{\csab}{$B\bot (ab)$}
\newcommand{\cpab}{$B\|(ab)$}
\newcommand{\tn}{$T_{\mathrm N}$}

\newcommand{\pbvo}{Pb$_2$VO(PO$_4$)$_2$}
\newcommand{\szvo}{SrZnVO(PO$_4$)$_2$}
\newcommand{\aavo}{AA'VO(PO$_4$)$_2$}
\newcommand{\Jo}{\ensuremath{J_1}}
\newcommand{\Jt}{\ensuremath{J_2}}

\begin{document}
\preprint{F}
\title{Spin fluctuations with two dimensional XY behavior in a frustrated S=1/2 square lattice ferromagnet}

\author{T. F\"{o}rster}
\author{F.A. Garcia}
\author{T. Gruner}
\author{E.E. Kaul}
\author{B. Schmidt}
\author{C. Geibel}
\author{J. Sichelschmidt}

\affiliation{Max Planck Institute for Chemical Physics of Solids, 01187 Dresden, Germany}

\date{\today}

\begin{abstract}
[Journal ref.: PRB 87, 180401(R) (2013)] The spin dynamics of the layered square-lattice vanadate \pbvo\ is investigated by electron spin resonance at various magnetic fields and at temperatures above magnetic ordering. 
The linewidth divergence towards low temperatures seems to agree with isotropic Heisenberg-type spin exchange
suggesting that the spin relaxation in this quasi-two dimensional compound is governed by low-dimensional quantum fluctuations. However, a weak easy-plane anisotropy of the $g$ factor points to the presence of a planar XY type of exchange.
Indeed, we found that the linewidth divergence is described best by XY-like spin fluctuations which requires a single parameter only. Therefore, ESR-probed spin dynamics could establish \pbvo\ as the first \textit{frustrated} square lattice system with XY-inherent spin topological fluctuations.
\end{abstract}

\maketitle

One of the central topics in magnetism of low-dimensional spin systems is the description of real magnetic systems by idealized models of magnetic interaction. Frustration effects and competing interactions often result in a critical behavior near transitions toward novel phases of matter which, for instance, could be described as spin ice \cite{Bram01} or quantum spin liquid \cite{Bale10} ground states. The critical spin dynamics was widely investigated by local spin probes with techniques like nuclear magnetic resonance (NMR) or electron spin resonance (ESR). In particular, for two-dimensional ($2D$) magnetic systems such investigations raised challenging questions to the applicability of spin models like the isotropic Heisenberg or the planar XY model.\cite{Benn90} 
The latter model is relevant for the recently often discussed topological spin excitations.
As shown by Kosterlitz and Thouless (KT) the $2D$ XY model has a topological phase transition of magnetic vortices to a state of vortex-antivortex pairs \cite{Kost73}. These magnetic vortices may sensitively impact the spin relaxation in the critical region as was shown by an analysis of the ESR line broadening for various quasi-$2D$ triangular and honeycomb lattices.\cite{Beck96,Hein03,Hemm09,Hemm11} \\
\indent In this context, the layered vanadium-oxide-bis(phosphates) \aavo\  present an interesting class of compounds. The crystal structure is dominated by layers of VO$_5$-pyramids and PO$_4$-tetrahedra forming a quasi-two dimensional (quasi-2$D$) $S=1/2$ \textit{square-lattice} of V$^{4+}$ ions \cite{Shpa06}. In contrast to the majority of 2$D$ square-lattice compounds, these vanadium systems present a significant amount of frustration because the superexchange coupling constant along the sides of the squares (\Jo) is of the same order of magnitude as the one along the diagonals (\Jt). Therefore, for systems like \pbvo\ and \szvo, magnetic susceptibility \cite{Kaul04} and neutron scattering data \cite{Skou09} were discussed in terms of the S=1/2 Heisenberg \Jo-\Jt-model on a square lattice \cite{Shan04,Schm07a}. 

Quantum Monte Carlo methods revealed a temperature driven crossover from an isotropic Heisenberg to a planar XY behavior if a weak easy-plane anisotropy or a magnetic field is present \cite{Cucc03,Cucc03b}. This crossover is predicted to occur around 30\% above the critical temperature $T_{\rm KT}$ and is reflected, for instance, in a developing anisotropy of the magnetic susceptibility. Interestingly, in both compounds \pbvo\ and \szvo\ properties of the $2D$-XY model could already be identified by NMR and $\mu$SR measurements in the spin dynamics above the long-range magnetic order at $T_{\rm N}$ and in the critical exponent of the order parameter  \cite{Carr09,Nath09,Boss11}. Noteworthy, we are aware of only three (unfrustrated) $S=1/2$ \textit {square-lattice} systems showing XY behavior, namely Sr$_2$CuO$_2$Cl$_2$ \cite{Cucc03}, Cu(pz)$_2$(ClO$_4$)$_2$ \cite{Tsyr10} and [Cu(pyz)$_2$(pyO)$_2$](PF$_6$)$_2$\cite{Koha11}. The reason for this scarcity is based on both the weakness of the easy-plane anisotropy and the existence of interlayer interactions in real quasi-2$D$ systems leading to a 3$D$ magnetic phase transition at a temperature slightly above the KT transition \cite{Bram93,Bram94}. Therefore, anomalies of the $3D$ transition easily mask XY anomalies which then appear too small in bulk properties like the specific heat.\\ 
\indent ESR has proven to be one of the most efficient methods to characterize the microscopic mechanism of the critical spin dynamics involving, in particular, magnetic vortices.The free movement of the vortices above the KT-transition increases the spin relaxation and leads to a decorrelation of the excited spins. This causes the ESR linewidth to be proportional to the inverse correlation length of the vortices \cite{Beck96,Hein03}. Here we show that the ESR line broadening observed for single crystalline \pbvo\ can be most reasonably explained by a XY behavior of the spin dynamics. This suggests this compound to be a rare case of a frustrated $S=1/2$ quantum spin system on a square lattice with a genuine XY behavior.\\
%
\begin{figure}[t]
  \centering
 \includegraphics[width=0.95\columnwidth]{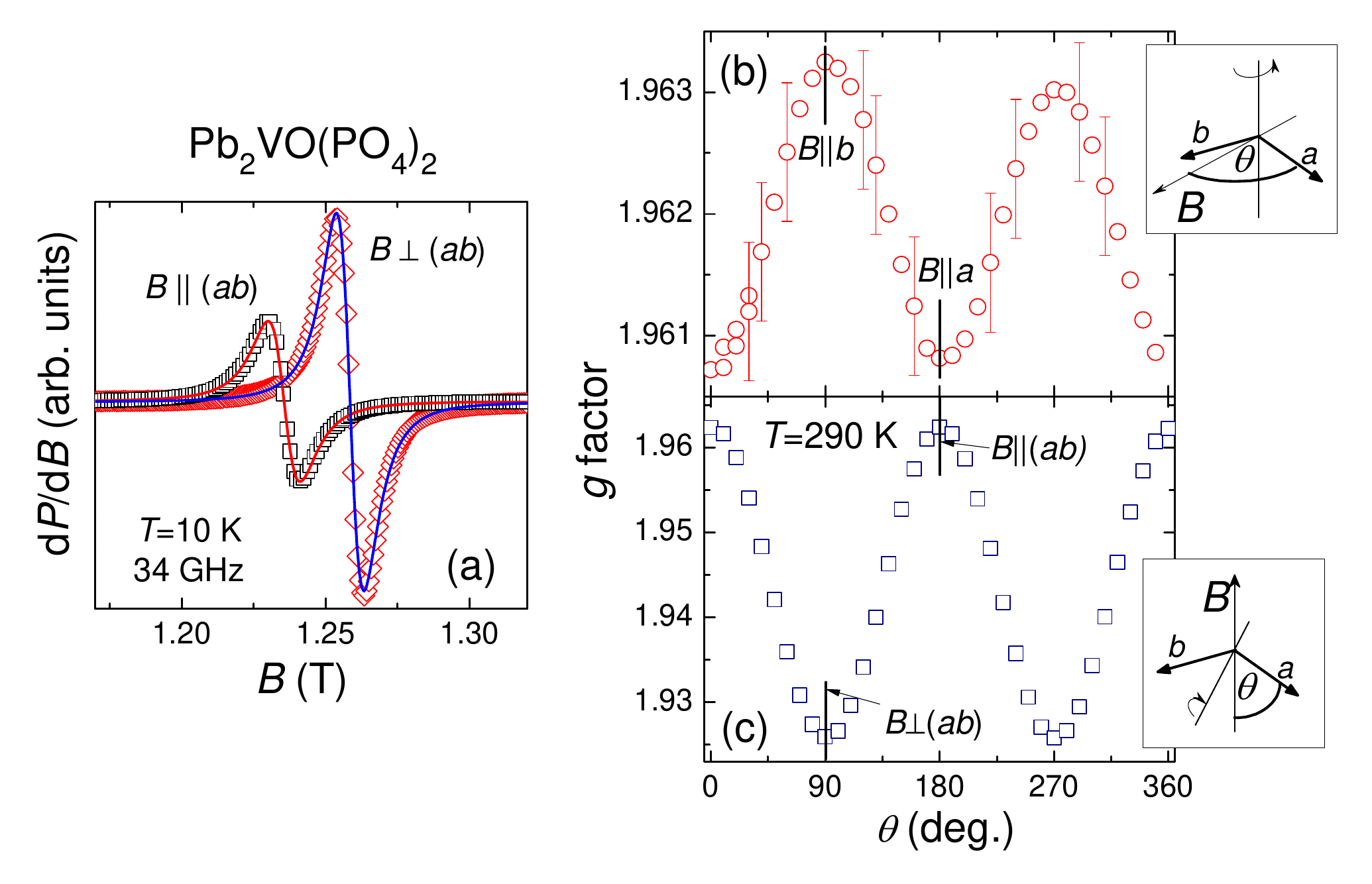}
  \caption{(a): typical ESR spectra of \pbvo\ at two crystal orientations in the external field $B$. Solid lines present Lorentzian lines. b,c): angular dependencies of ESR $g$-factor at 34~GHz and $T=290$~K for two geometries of crystal rotation by an angle $\theta$ as defined in the sketches: rotation around an axis perpendicular to $(ab)$-plane (frame b), and around the [110]-axis  (frame c). Note the different scales of the frames.}
 \label{signal+g-anisotropy}
\end{figure}
%
\indent The synthesis of the investigated single crystals of \pbvo\  (monoclinic structure, $P2_1/a$) was reported in Ref. \cite{Shpa06}. The dark-green samples are transparent, reflecting the electronic band gap of an insulator. Using data of susceptibility and neutron scattering the strength of the exchange interactions could be estimated as $J_1/k_{\mathrm B}= -3.2$~K (ferromagnetic) and $J_2/k_{\mathrm B} = 7.7$~K (antiferromagnetic)\cite{Skou09}. The interlayer exchange induces a columnar antiferromagnetic order at \tn$=3.5$~K \cite{Kaul05}. The specific heat $C(T)$ was determined with a PPMS using a relaxation method, while the susceptibility $\chi(T)$ was measured in a SQUID.\\
\indent The ESR experiments were carried out using a continuous wave spectrometer together with He-flow cryostats allowing for temperatures between 1.5 and 300~K. ESR probes the absorbed power $P$ of a transversal magnetic microwave field at frequency $\nu$ as a function of a static magnetic field $B$. To improve the signal- to-noise ratio, a lock-in technique yielded the derivative of the resonance signal $dP/dB$. At all temperatures above \tn\ we observed very well defined signals which could be accurately fitted with single Lorentzian lines (solid lines in Fig.\ref{signal+g-anisotropy}a). The relatively small linewidths allowed a precise determination of the ESR parameters linewidth $\Delta B$ and resonance field $B_{\mathrm{res}}$. From the resonance condition $g=h\nu/\mu_{\mathrm B} B_{\mathrm{res}}$ one obtains the V$^{4+}$ $g$ factors, which agree well with V$^{4+}$ $g$ factors for compounds with similar VO$_5$-pyramids like Sr$_2$V$_3$O$_9$ \cite{Ivan03} and NaV$_2$O$_5$ \cite{Lohm00a}.\\
%
\begin{figure}[t]
  \centering
 \includegraphics[width=0.9\columnwidth]{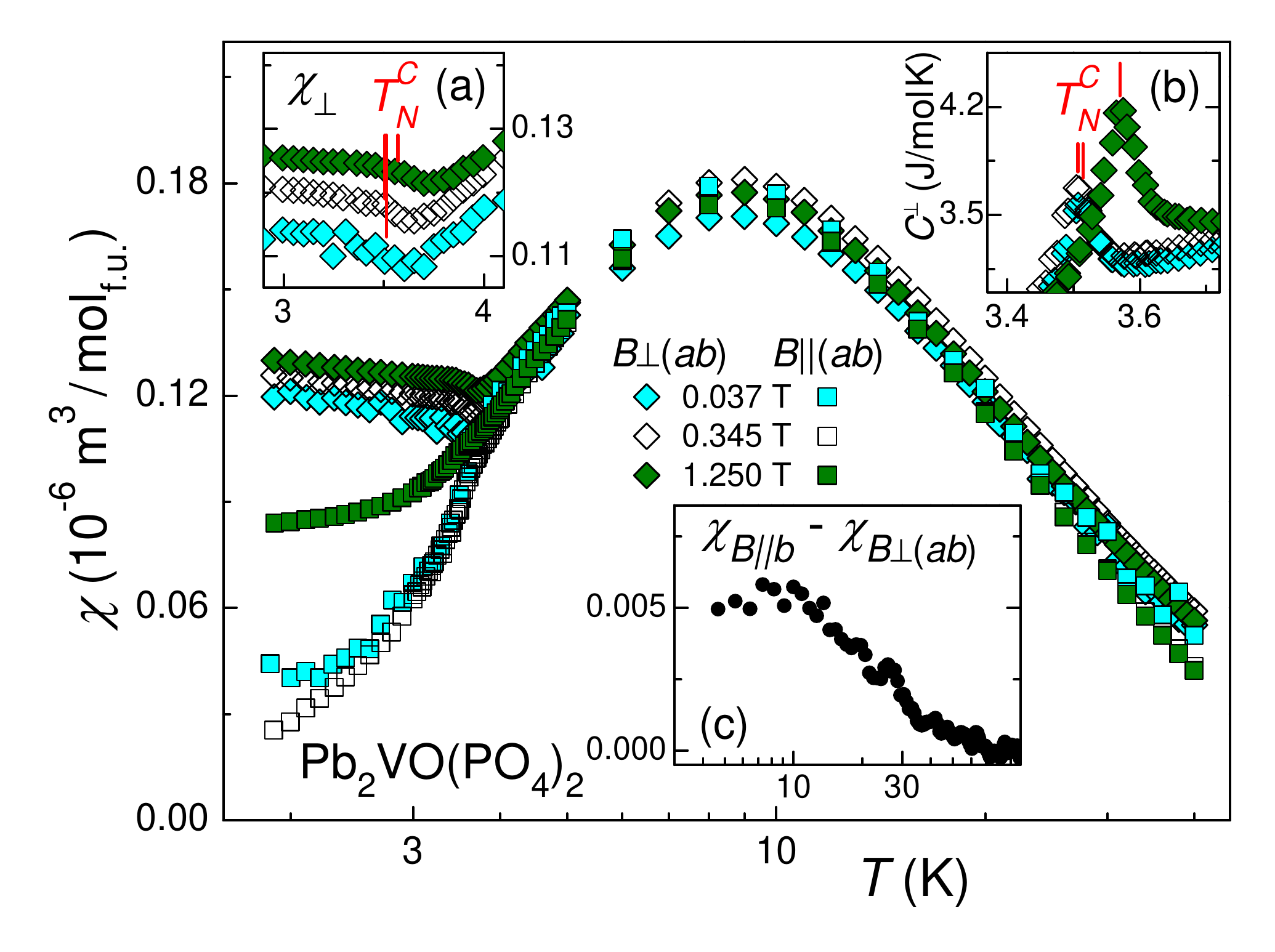}
  \caption{Magnetic susceptibility $\chi(T)$ for $B\|(ab)$ and $B\bot (ab)$ at different fields corresponding to $B_{\mathrm{res}}$ at the used frequencies. Inset (a): $\chi_{B\bot (ab)}$ with red lines indicating $T^{C_{\bot}}_{\mathrm{N}}$ obtained from specific heat measurements at the same fields shown in inset (b). Inset (c) presents anisotropy of $\chi(T)$ at $B=0.1$~T of a large single crystal of \pbvo\  \cite{Kaul05}.}
 \label{susc}
\end{figure}
%
\indent The $g$ factors sensitively depend on the crystal orientation in the external field. Besides a tiny in-plane anisotropy (\autoref{signal+g-anisotropy}b), they display even at $T=290$~K a weak easy-plane anisotropy of $\approx2$\% (\autoref{signal+g-anisotropy}c) which is remarkably small compared to other XY-type systems like [Cu(pyz)$_2$(pyO)$_2$](PF$_6$)$_2$ \cite{Koha11}. This anisotropy is also reflected in the magnetic susceptibility $\chi(T)$ as plotted in \autoref{susc}   which we measured for the three ESR resonance fields. Interestingly this is already visible at quite high temperatures well above 30~K (see inset (c) of \autoref{susc}). As shown in inset (a) of \autoref{susc} the transition temperature $T^{C_{\bot}}_{\mathrm{N}}$ ($B\bot (ab)$), obtained from specific heat measurements and indicated by thin bars, is located slightly below a kink in transverse $\chi_{\bot}(T)$. This observation is a clear signature of a dimensional crossover \cite{Cucc03,Cucc03b}, thus providing another important requirement that, indeed, for \pbvo\ the spin dynamics may be characterized by a crossover from an isotropic (Heisenberg) to a planar XY spin model.\\
%
\begin{figure*}[t]
  \centering
 \includegraphics[width=0.72\textwidth]{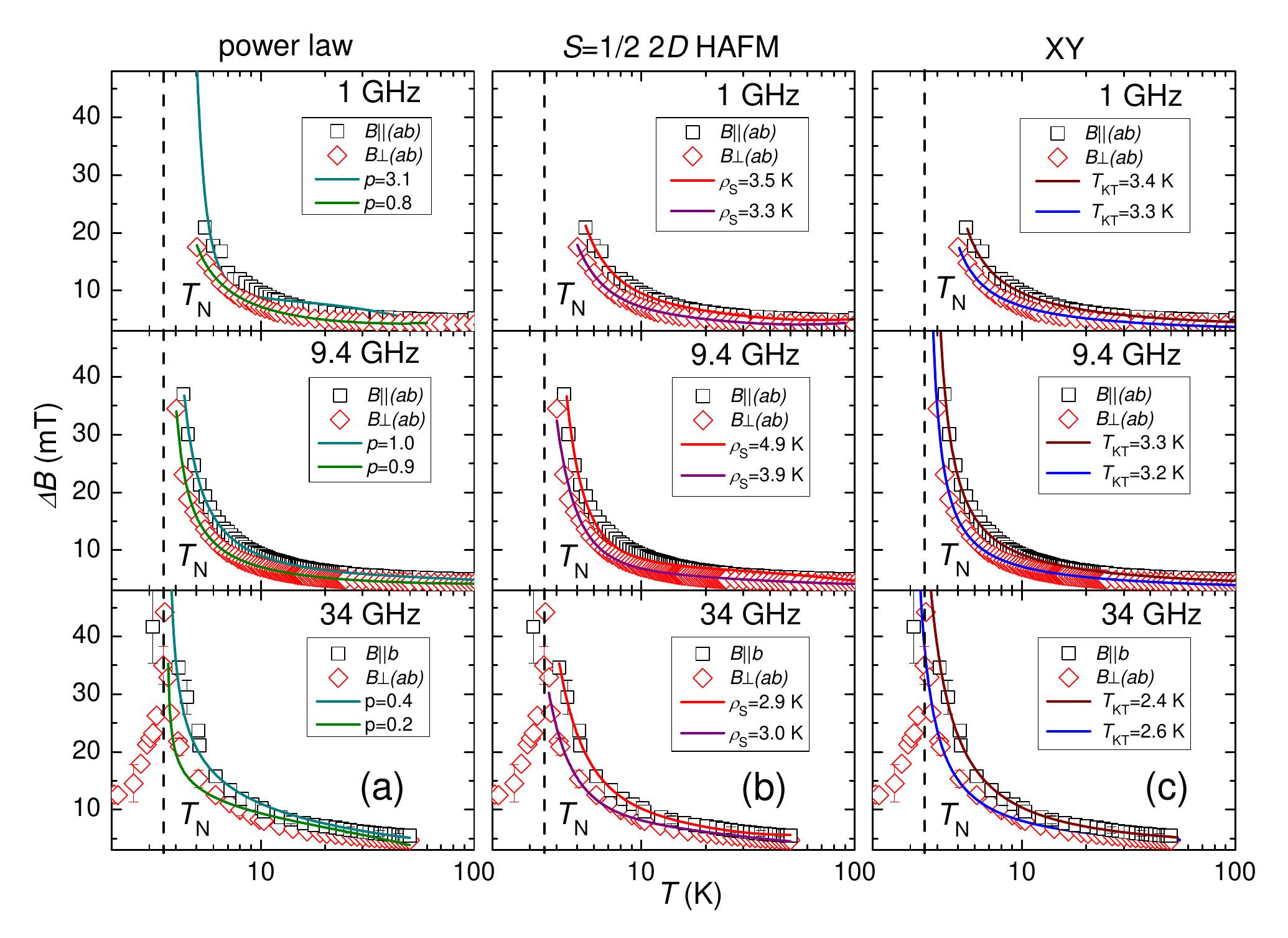}
  \caption{Temperature dependence of linewidth $\Delta B$ of \pbvo\ at different orientations of the magnetic field. The solid lines represent data fits with $\Delta B_{0}+\Delta B^{\mathrm{crit}}$ with $\Delta B_{0}=bT+b_{0}$ and $\Delta B^{\mathrm{crit}}$ according equations \ref{eq:powerlaw} (frame a), \ref{eq:2DHAFM} (frame b), and \ref{eq:KT-Transition} (frame c). The XY model (frame c) could fit the data without any non-critical contribution $\Delta B_0$.}
 \label{Fits}
\end{figure*}
%
\indent In ESR the spin dynamics is accessible by investigating the spin relaxation rate which determines the linewidth of the ESR spectra. \autoref{Fits} shows the temperature behavior of the ESR linewidth for the three used excitation frequencies 1.0, 9.4 and 34~GHz and for parallel and perpendicular orientation of the $(ab)$-plane in the magnetic field $B$. At temperatures $T>100$~K the effective linewidth is determined by an exchange narrowed dipolar broadening \cite{Ande53}. Using the energy scale of the exchange interaction $J_c/k_{\mathrm B}=\sqrt{J_{1}^{2}+J_{2}^{2}}/k_{\mathrm B}=8.3$~K \cite{Skou09} and a linewidth due to dipole-dipole interaction of $\approx 100$~mT we estimate an effective linewidth of $\approx3$~mT which roughly agrees with the observed high-temperature limit of (4.1$\pm$0.3)~mT for \csab\ and (4.9$\pm$0.3)~mT for \cpab. By lowering the temperature towards \tn\ the critical spin dynamics leads to a divergent linewidth behavior. We analyzed this divergency with fitting the data by $\Delta B_{0}+\Delta B^{\mathrm{crit}}$, i.e. by a part $\Delta B_0(T)=bT+b_{0}$ containing all non-critical contributions and by a critical part $\Delta B^{\mathrm{crit}}$ given by the following models:\\
\indent The ``power law'' model (\autoref{Fits}, frame a) considers classical critical fluctuations of a $3D$ order parameter and is based on a suppression of the above discussed exchange narrowing by 3D order parameter fluctuations. These extend well into the paramagnetic region up to 20~K as indicated by diffuse neutron scattering data \cite{Skou09} and lead to the following linewidth divergence with a critical exponent $p$ \cite{Benn90}:
\begin{equation}
\Delta B_{\mathrm P}^{\mathrm{crit}} = C_1 \left( \frac{T}{T_{\mathrm N}}-1\right)^{-p}
\label{eq:powerlaw}
\end{equation}
Here, $C_1$ is a temperature independent parameter. The data fitting with $T_{\mathrm N}=3.5$~K results in a rather poor agreement for the 1 and 34~GHz data and, moreover, in wide spreaded values of $p$. The fit qualities could not be improved by using the $T_{\mathrm N}$ values we obtained from the field- and orientation dependent specific heat data, i.e. by using values $3.4-3.6$~K. Given this disparity we conclude that critical fluctuations of the 3$D$ order parameter cannot describe the ESR linewidth divergency above \tn.\\
\indent A description of the ESR linewidth divergency within a $S=1/2$ $2D$ antiferromagnetic Heisenberg model ($2D$ HAFM) on the square lattice was first discussed for the spin relaxation in La$ _2$CuO$_4$ \cite{Chak90}. In this model the linewidth depends on the spin correlation length $\xi$ of the $2D$ HAFM in relation to the nearest neighbor distance of the spins on the square lattice $d$
\begin{equation}
\frac{\xi}{d} = \xi_0 \frac{\exp{\left( 2\pi \rho_s/k_{\mathrm B}T \right)}}{1+k_{\mathrm B}T/ 2\pi
\rho_s}
\label{eq:2DCorrle}
\end{equation}
where $\rho_{\text s}$ is the spin stiffness.  Eq.\ref{eq:2DCorrle} is valid if $\pi\rho_{\text s}\ge k_{\text B}T$~[27].  According to~\cite{halperin:69}, for an antiferromagnet the spin stiffness is given by $\rho_{\text s}=\chi_{\perp}v_{\text s}^{2}$, where $\chi_{\perp}$ is the uniform susceptibility in units such that $(N/V)\mu_{0}(g\mu_{\text B})^{2}=1$, and $v_{\text s}$ is the spin wave velocity. Omitting higher-order corrections, we get from linear-spin wave theory
\begin{equation}
    \rho_{\text s}^{(\alpha)}
    \approx
    4S^{2}\chi_{\perp}
    \times
    \left\{
    \begin{array}{c@{,\ }c}
        \left(2J_{2}+J_{1}\right)^{2} & \alpha=x
        \\
        2J_{2}^{2}-J_{1}^{2} & \alpha=y
    \end{array}
    \right\}
    \times
    \left(1-\eta^{2}\right)
\end{equation}
where $\alpha=x,y$ denote the two directions parallel and perpendicular to the ordering vector $\vec Q$.  Here $\eta=H/H_{\text{sat}}$ parametrizes the dependence on a small applied field $\mu_{0}H\ll\mu_{0}H_{\text{sat}}=4S(J_{1}+2J_{2})/(g\mu_{\text B})\approx20.9\,\text T$~\cite{tsirlin:09b}.  With $\eta\lesssim0.06$ in our experiments, we will neglect this small correction.  Within our approximation, the transverse susceptibility is given by its classical value $\chi_{\perp}=1/(4(2J_{2}+J_{1}))$.
Putting in numbers yields $\rho_{\text s}^{x}/k_{\text B}=3.05\,\text K$, and $\rho_{\text s}^{y}/k_{\text B}=4.65\,\text K$. From our analysis, we get an average value $\rho_{\text s}/k_{\text B}=3.7\,\text K$.
Thus, we may use Eq.\ref{eq:2DCorrle} up to $T\le 11.6$~K for the critical line broadening which is calculated as
\begin{equation}
\Delta B_{\mathrm{HAF}}^{\mathrm{crit}} = C_2 \left(\frac{\xi}{d} \right)^3 \frac{\left(k_{\mathrm B}T/ 2\pi \rho_s \right)^{5/2}}{\left
( 1+ k_{\mathrm B}T/ 2\pi \rho_s\right)^4}
\label{eq:2DHAFM}
\end{equation}
where the temperature independent $C_2$ contains renormalizations of $\chi$ due to quantum fluctuations \cite{Chak90}. This model describes the data fairly well as shown in \autoref{Fits}b by values of $\rho_s$ showing reasonable agreement with theoretical $\rho_s$ in the \Jo-\Jt-model \cite{Shan04}. However, it was impossible to describe the data set with a common (i.e. field independent) $\rho_s$ rendering the $2D$ HAF model for $\Delta B^{\mathrm{crit}}$ questionable. Moreover, as in the case of power law fitting, the validity of this model is restricted by the requirement of a non-critical contribution $\Delta B_{0}$.

The movement of the vortices in the XY model leads to a spin relaxation which depends on the density of these vortices. Neglecting a temperature dependence of the vortex velocity one finds \cite{Beck96,Hemm09}
\begin{equation}
\Delta B_{\mathrm{XY}}^{\mathrm{crit}} = C_3 \exp{\left( \frac{3\beta}{\sqrt{\frac{T}{T_{\mathrm{KT}}}-1}}\right)} 
\label{eq:KT-Transition}
\end{equation}
assuming that the contribution of out-of plane correlations of the vortices is very small and thus can be neglected \cite{Kost74}. $T_{\mathrm{KT}}$ is the critical temperature of the KT transition and $C_3$ is constant. For the $\beta$-parameter experimental values lie between 0.9 (NMR,\cite{Gave91}) and 4.1 (ESR,\cite{Hemm09}) in agreement with theoretical estimations \cite{Kost74,Mert89}. 
We note that \pbvo\ is \textit {not} a Kosterlitz-Thouless system.  A KT transition, being a very special $2D$ phenomenon, cannot occur here. In particular, due to the dimensional crossover $3D$ N\'eel order sets in at $T=T_{\text N}>T_{\text{KT}}$ (see our analysis below).  Instead, $T_{\text{KT}}$ is to be understood here as an energy scale of the $2D$ XY type spin vortex fluctuations occuring at $T>T_{\text N}$.

The data description with \autoref{eq:KT-Transition} are presented in \autoref{Fits}c showing a very good agreement. In contrast to the other models \autoref{eq:KT-Transition} describes the data without needing a non-critical term $\Delta B_0(T)$, reducing the number of the free fit parameters to three only. It is worth to note that the $T_{\mathrm{KT}}$ values (displayed in \autoref{Fits}c) remain almost unchanged even if the above estimated high-temperature limit $b_{0}=3$~mT is added to  \autoref{eq:KT-Transition}. The extraordinary fit quality is furthermore supported by the stability of the exponent 0.5 in the denominator of \autoref{eq:KT-Transition} as soon as it is assumed to be a free parameter. The parameters $C_{3}$ and $\beta$ show only moderate variations with values in the theoretically expected range \cite{Mert89}. The obtained temperatures of the KT transition are smaller than \tn\ and are in good agreement with $T_{\mathrm{KT}}\approx0.85\,$\tn\ as estimated according to Ref. \cite{Bram93}. We relate the continuous decrease of $T_{\mathrm{KT}}$ upon increasing the resonance field (0.037, 0.345, and 1.25 T at frequencies 1, 9.4, and 34~GHz) to a decrease of the vortex-binding energy with increasing field. A comparison of the field behavior of $T_{\mathrm{KT}}$ with phase diagrams for field-induced XY-behavior \cite{Cucc03b} is not meaningful because these diagrams are based on a model without frustration. We are not aware of any theoretical approach treating the vortex-binding energy in a \textit{frustrated} \Jo-\Jt-XY model (which is applicable for \pbvo).

An in-plane field is supposed to break the rotational symmetry necessary for XY behavior. However, for our system and field values the model appears to be stable and applicable for both field directions. The stability of the XY behavior may be related to the frustration in the system. A similar behavior is found in the frustrated square lattice Cu(pz)$_2$(ClO$_4$)$_2$ \cite{Tsyr10}. We expect that our magnetic fields are not large enough to suppress the XY behavior in \pbvo. A negligible influence of small in-plane fields on the vortex dynamics was theoretically discussed for systems with easy-plane anisotropy \cite{Pere95}.\\
\indent Considering the temperature range of  validity for the XY model one notes that the question whether $2D$ spin dynamics can be referred to unbound vortices well above $T_{\mathrm{KT}}$ was often discussed in literature. For example, in-plane correlations may be characterized by a gas of vortices \cite{Mert89} and $2D$ topological spin vortices appear as preferable thermodynamic configurations to understand specific heat results  \cite{Tsyr10}. Also for ESR-probed spin dynamics one finds several examples confirming divergent ESR linewidth behavior even far above $T_{\mathrm{KT}}$: For honeycomb and triangular compounds \cite{Hein03,Hemm09} it was shown that Eq. \ref{eq:KT-Transition} is valid well above $T_{\mathrm KT}$ (up to $T=3T_{\mathrm{KT}}$ (honeycomb), up to $T=30T_{\mathrm{KT}}$ (triangular)). Especially for frustrated systems (like \pbvo) the KT-critical region is observed and predicted to be large \cite{Hemm11}.\\
\indent The good agreement of both the 2$D$-HAF and XY models with the data establishes 2$D$ spin fluctuations to be responsible for the ESR linewidth divergency. The dimension of spin fluctuations, however, is best described by a planar XY model, which renders a more consistent physical picture for the behavior of the fitting parameters. This is furthermore supported by the observed susceptibility signatures (\autoref{susc}(a)) of a dimensional crossover in spin space.\\
\indent To summarize, we investigated the spin dynamics of the quasi-2$D$ square lattice compound \pbvo\ by analyzing the ESR linewidth with various spin interaction models. 
For this compound magnetic susceptibility shows the hallmarks for a crossover from Heisenberg to XY behavior as predicted by quantum Monte-Carlo calculations \cite{Cucc03b}. We found clear support for XY behavior of the spin correlations according the analysis of the ESR linewidth divergency together with the weak in-plane anisotropy in the ESR $g$ factor. XY-type spin dynamics could clearly be verified in the spin relaxation, thus providing a new basis for understanding frustrated square-lattice systems. These findings confirm previous indications of XY behavior observed by NMR and $\mu$SR measurements in this and related compounds \cite{Carr09,Nath09,Boss11}. However, we have shown that measuring ESR linewidth and in-plane-anisotropy in \pbvo\ is accurate enough to clearly set apart the XY spin dynamics model from other models, hence establishing \pbvo\ to be one of the rare cases of a $S=1/2$ spin quantum system showing a crossover from Heisenberg to XY symmetry.
\begin{acknowledgments}
We acknowledge fruitful discussions with H.-A. Krug von Nidda (Augsburg) and M. Hemmida (Augsburg).
\end{acknowledgments}
%
%

\end{document}